\documentclass[11pt,journal,letterpaper,twoside,onecolumn,draftcls]{IEEEtran}
\usepackage{trans-sp-2015}

\title{Interactive Scalar Quantization for Distributed Resource Allocation}
\author{%
Bradford~D.~Boyle,~\IEEEmembership{Student~Member,~IEEE,} %
Jie~Ren,~\IEEEmembership{Student~Member,~IEEE,} %
John~MacLaren~Walsh,~\IEEEmembership{Member,~IEEE}, and %
Steven~Weber,~\IEEEmembership{Senior~Member,~IEEE}%
\thanks{%
This research has been supported by the Air Force Research Laboratory under agreement number FA9550-12-1-0086. %
The U.S.\ Government is authorized to reproduce and distribute reprints for Governmental purposes notwithstanding any copyright notation thereon.
}
\thanks{%
The authors are with the Department of Electrical and Computer Engineering, Drexel University, Philadelphia, PA USA. %
The contact author is Steven Weber (email: \textsf{sweber@coe.drexel.edu}).%
}}

\begin{document}
	\maketitle
	\begin{abstract}
		In many resource allocation problems, a centralized controller needs to award some resource to a user selected from a collection of distributed users with the goal of maximizing the utility the user would receive from the resource.
		This can be modeled as the controller computing an extremum of the distributed users' utilities.
		The overhead rate necessary to enable the controller to reproduce the users' local state can be prohibitively high.
		An approach to reduce this overhead is interactive communication wherein rate savings are achieved by tolerating an increase in delay.
		In this paper, we consider the design of a simple achievable scheme based on successive refinements of scalar quantization at each user.
		The optimal quantization policy is computed via a dynamic program and we demonstrate that tolerating a small increase in delay can yield significant rate savings.
		We then consider two simpler quantization policies to investigate the scaling properties of the rate-delay trade-offs.
		Using a combination of these simpler policies, the performance of the optimal policy can be closely approximated with lower computational costs.
	\end{abstract}

	\section{Introduction}
	A common pattern in resource-limited systems is the allocation of the resource by a controller among a set of competing consumers/users.
	In an effort to make the most efficient usage of the resource, the controller awards the resource to the user that derives the most utility from it.
	This can be modeled as the controller computing an extremum (max, min, arg max, arg min) of the distributed users' utilities.
	By computing the \(\argmax\), the controller can allocate the resource to an appropriate user without knowing exactly its value.
	Computing the \(\max\) of the users' local state allows the controller to make a global decision, e.g., turning on building-wide air conditioning based on the hottest room.
	Computing both the max and the arg max across users would provide the controller with knowledge of both the user to award the resource to, and the utility this best user would receive from it.

	Often, these users are not colocated and must communicate their utility to the centralized controller.
	This communication presents an overhead and techniques for minimizing the required rate are needed.
	In some cases, the rate can be reduced by incurring only a small penalty.
	Rate-distortion theory is an example where the rate required to communicate/reconstruct a signal can be reduced if small errors in the reconstruction are tolerated.
	Interactive communication allows an alternative framework wherein communicating parties can send messages back-and-forth over multiple rounds \cite{Kas1985}.
	This back-and-forth messaging can reduce the rate required to compute an extremum/extrema of the sources at the cost of increased delay.
	In the rate-distortion framework, an achievable scheme for a collection of distributed users talking to central controller (multiterminal CEO) is quantization followed by entropy (i.e., Huffman) encoders \cite{FlyGra1987,PraKusRam2002}.
	In this paper, we consider this same technique in the context of interactive communication.
	We formulate the design of the multi-round quantization as the solution to a dynamic program and demonstrate a substantial reduction in overhead communication rate can be obtained for a small increase in delay.
	The use of dynamic programming to solve for optimal quantization policies in the interactive multiterminal source coding problem parallels earlier work that used dynamic programming to solve for optimal quantization of a single source \cite{Sha1978}.
	Because of the high computational complexity of the dynamic programming problem, we identify two simpler schemes and investigate their performance over a range of parameters.
	By combining these two schemes, we demonstrate a close approximation of the dynamic programming solution at a lower computational cost.

	The paper is organized as follows.
	We review related literature from information theory, signal processing, and communication complexity in \secref{related-work}.
	We present the basic problem model and establish the mathematical notation used throughout the paper in \secref{problem-model}.
	In \secref{dyn-prog}, we formulate the optimal rate-delay trade-off of scalar quantization as the solution of a minimum cost dynamic program.
	We characterize the set of terminating states for the dynamic program when computing the different extremum functions under consideration (cf.\ \propref{argmax-stop} \& \propref{max-stop}).
	We prove that, asymptotically in the number of users, the cost for the CEO to compute the different extremum functions are equal (cf.\ \propref{max-upper-bound}).
	In \secref{simple-schemes}, we restrict our attention to the case of users' utilities being distributed uniformly.
	This assumption allows us to provide analytical expressions for the rate and delay of simple quantization policies (cf.\ \propref{binary-search} \& \propref{max-search}).
	We then provide an extension to these simple schemes, which is asymptotically (in the number of users) sufficient for minimizing the cost of computing the selected extremum functions (cf.\ \propref{asymptotically-optimal}).
	The proposed family of quantizers is significantly smaller compared to the space of all possible quantizers.
	Hence, with these results the dynamic program can be solved more quickly with a relatively small incurred penalty relative to the optimal dynamic program.
	In \secref{results}, we show the rate-delay trade-offs for different distributions and varying numbers of users.
	We also present a comparison of the minimum cost from searching over all quantizers to the minimal cost from searching over our proposed family of quantizers.
	We present directions for future work in \secref{conclusions}.

	\section{Related Work}
	\label{sec:related-work}
	In the present work, we consider a distributed model where all users communicate to a central node that wishes to compute a function of the users' sources.
	In particular, the users cannot overhear each other but can design their codes (with knowledge of the source distributions of other users) to communicate with the central node---called ``cooperative design, separate encoding'' \cite{LonLooGra1990}.
	These models are referred to as the chief estimation officer (CEO) problem \cite{BerZhaVis1996}.
	We begin by reviewing fundamental limits and achievable schemes for non-interactive variants of CEO-type problems.
	We then review results for interactive variants that demonstrate significant rate savings may be possible.
	The non-interactive CEO problem has received considerably more attention than the interactive variant, and, for the cases where fundamental limits are known,  quantization followed by entropy coding closely approximates these limits.
	This motivates our study of interactive quantization as a means to realize further rate savings at the expense of an incurred delay.

	\subsection{Non-interactive communication: fundamental limits}
	In information theory, the interest is usually on characterizing inner/outer bounds for the rate region.
	Berger et al.\ introduced the generic CEO problem, wherein the CEO wants to reproduce the source from the received signals \cite{BerZhaVis1996}.
	The rate region for the problem of source reproduction with constrained distortion remains unknown, except for the cases of:
	\begin{inparaenum}[i)]
		\item jointly Gaussian sources with quadratic distortion \cite{WagAna2008,PraTseRam2004, Ooh2005};
		\item finite alphabet sources with logarithmic loss \cite{CouWei2014}; or,
		\item discrete source distributions which are independent across users \cite{KuRenWal2015}.
	\end{inparaenum}
	Vempaty and Varshney considered the CEO problem for non-regular source distributions (e.g., truncated Gaussian) with quadratic distortion and studied the asymptotic behavior of the distortion function \cite{VemVar2015}.

	In our work, we are interested in having the CEO compute a function of all sources; this is referred to as the \emph{distributed function computation (DFC) problem}, and was considered in \cite{HanKob1987,SefTch2013,DosShaMed2010}.
	This general formulation contains the specialized problem of function computation with side information; in this problem, the CEO knows all but one of the sources \cite{Wit1976,OrlRoc2001}.
	When the problem requires error-free computation, it was shown that the minimum worst-case rate is related to the chromatic number of the characteristic graph of the source \cite{Wit1976}.
	In the case of lossless (in the Shannon sense) computation, it was shown that the minimum average rate is the conditional graph entropy of the characteristic graph of the source \cite{OrlRoc2001,Kor1973}.
	Building upon this line of research, the rate region for the lossless DFC problem was characterized for certain problem instances \cite{SefTch2013,DosShaMed2010}.
	Sefidgaran et al.\ derived inner and outer bounds to the rate region for a class of tree-structured networks (which includes the CEO problem) and showed that the inner and outer bound coincide with each other if the sources obey certain Markov properties \cite{SefTch2013}.
	Doshi et al.\ gave the rate region for the DFC problem under a different constraint that they referred to as the ``zig-zag" condition \cite{DosShaMed2010}.
	They showed that any achievable rate point can be realized by graph coloring at each user and Slepian-Wolf (SW) \cite{SleWol1973} encoding the colors.
	Han and Kobayashi partitioned all DFC problems based on whether their achievable rate region coincides with the SW region \cite{HanKob1987}.

	The aforementioned literature provides insightful outer bounds for comparing the performance of distributed quantizer designs \cite{BoyWalWeb2014,RenBoyKu2014}, but the achievable schemes used in the proofs usually require block coding with infinite block length, which is not practical.
	For use in a real system, simpler achievable schemes with low computational complexity and performance close to the limits are needed.

	\subsection{Non-interactive communication: achievability}
	A concern of signal processing is to provide optimized practical quantization algorithms for the DFC system with performance close to the rate-distortion limits \cite{FlyGra1987,PraKusRam2002,LonLooGra1990,Gra2006,MisGoyVar2011,SunMisGoy2013,FleZhaEff2004}.
	There are asymptotic results for sufficiently high-rate and low-distortion that are derived by applying high rate quantization theory \cite{LooGra1989}, while there are also non-asymptotic results derived from generalizations of Lloyd's algorithm.

	For the high-rate and low-distortion scenario, Misra et al.\ considered a quantization scheme for the analysis of distributed scalar quantization \cite{MisGoyVar2011}.
	It was shown that, with certain constraints on the objective function and source distributions, the high-resolution approach can asymptotically achieve the rate-distortion limits, and the optimized quantization is regular\footnote{A quantizer is called regular if each partition cell is an interval and each output level is within the corresponding interval.}.
	Sun et al.\ used a similar high-resolution approach, but with a simpler decoder design and relaxed source distribution requirements \cite{SunMisGoy2013}.

	For the general rate-distortion problem, an algorithm for building optimized distributed quantizers was given wherein the CEO uses the quantized observations to perform hypothesis testing \cite{LonLooGra1990}.
	A two-stage distributed scheme was proposed for the case when the users each have a noisy observation on the same source, and the CEO needs to reproduce the source with a bounded expected distortion \cite{FlyGra1987,PraKusRam2002}.
	A first stage of local quantization is followed by a second stage of encoding the quantized signals based on Slepian-Wolf coding using syndrome codes \cite{PraKusRam2002} or index reuse techniques \cite{FlyGra1987}.
	In their treatment of the CEO problem with non-regular source distributions, Vempaty and Varshney provided an acheivability proof that utilized a layered approach of quantization followed by entropy coding \cite{VemVar2015}.

	Most of the provided distributed quantization schemes are non-interactive, which means the users each communicate with the CEO once, and no feedback is allowed from the CEO.
	For the problems in which the non-interactive fundamental limits are known, distributed quantization shows satisfactory performance as compared with the limit.
	However, very little work has been done in the interactive distributed cases.

	\subsection{Interactive communication}
	\emph{Interactive communication} is a scheme that allows message passing over multiple rounds.
	At each round, the communicating parties are allowed to send messages based on what they have received in previous rounds as well as their local source observation \cite{Kas1985}.
	The interactive communication literature is roughly divided into two categories: communication complexity and interactive information theory.

	The communication complexity literature is concerned with finding communication protocols that minimize the sum-rate subject to different sets of constraints.
	Overviews of communication complexity can be found in \cite{KusNis1997,Lov1990}.
	Communication complexity is defined as the sum-rate cost minimized over all protocols and maximized over all possible input pairs (worst-case cost).
	Average cost has also been studied for randomized coding protocols.
	Much of the communication complexity literature is focused on 2 users.
	Models with an \(N\)-terminal setup were considered in \cite{ChaFurLip1983,BabNisSze1992}, where the authors focused on providing communication complexity bounds with the restrictions that the function must be Boolean and the message sent at each round must be binary.
	However, in our work, we are interested in providing achievable schemes for problems without the limitation to 1 bit of communication in each round or the restriction to Boolean functions.

	Kaspi determined the two party information theoretic limit for lossy compression via interactive communication \cite{Kas1985}.
	This line of research was continued by Ma and Ishwar, who showed (by an example) that the minimum rate for a given distortion constraint can be arbitrarily smaller than the non-interactive minimum rate obtaining the same distortion \cite{MaIsh2010}.
	In follow up work, Ma and Ishwar showed (by an example) that for the DFC problem, the minimum sum-rate for losslessly computing a function can be smaller than the non-interactive rate; even infinitely-many rounds of interaction may still improve the rate-region \cite{MaIsh2011}.

	These results motivate us to consider interaction for the DFC problem.
	In earlier work, we considered the non-interactive DFC problem of computing an extremum of independent users.
	We developed distributed scalar quantizers with rate-distortion performance close to the rate-distortion limits \cite{BoyWalWeb2014,RenBoyKu2014}.
	We provided an achievable interactive communication scheme where the CEO communicates a threshold to the users at each round and the users reply with a single bit indicating if its value is above or below the threshold \cite{RenBoyKu2014,RenWal2014}.
	This scheme can be thought of as a simple two-bin quantizer selected by the CEO at the beginning of each round; in the present work we extend this by allowing the CEO to select a multi-bin quantizer in each round.

	This interactive coding scheme can be understood as a type of posterior matching \cite{ShaFed2011}.
	Shayevitz and Feder considered the problem of point-to-point communication over a memoryless channel with noiseless feedback from the receiver.
	A capacity-achieving transmission scheme was developed based on the transmitter providing statistically independent information that is missing by observing the a-posteriori density of the message as feedback from the receiver.
	In the present work, the focus is on minimizing the sum-rate from a collection of sources; however, the feedback from the CEO is used by the users in determining what is transmitted in the next round.

	\section{Problem Model}
	\label{sec:problem-model}
	We assume that we have \(N\) users, each with local state \(X_i\) for \(i = 1, \ldots, N\), vying for a resource to be allocated by the controller (i.e., CEO).
	We model the users' local states as i.i.d.\ (across users) discrete random variables with support set \(\mathcal{X}\) and PMF \(p_X(x)\), and interpret state as a proxy for the users' utility.
	Maximum utility of the resource is obtained by allocating it to the user with the largest local state value.
	Without loss of generality, we will take \(\mathcal{X} = \{1, \ldots, L\}\) where \(L\) is the size of support set.
	To select a user to award the resource to, the CEO wishes to compute one of the following functions\footnote{Though we focus exclusively on the case of maximization in this paper, similar results hold for the case of minimization. In later sections, we enforce a decreasing order on certain parameters; to obtain results for minimization, the order should be increasing.}:
	\begin{inparaenum}
		\item \(\argmax_i X_i\);
		\item \(\max_i X_i\), or;
		\item \((\argmax_i X_i, \max_i X_i)\).
	\end{inparaenum}
	\begin{figure}
		\centering
		\includegraphics{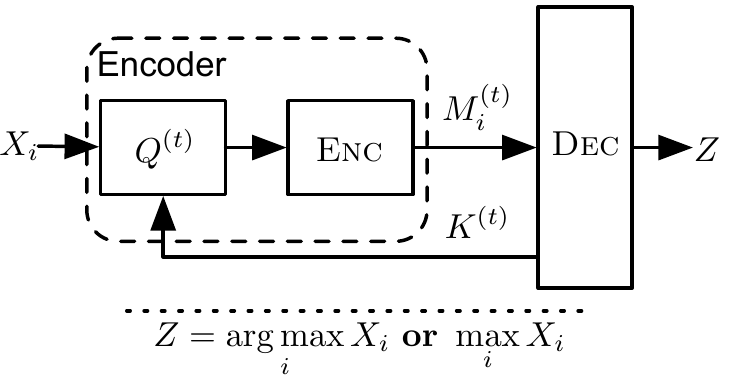}
		\caption{Interactive quantization system diagram. The users' utilities \(X_i\) are quantized using the quantization function \(Q^{(t)}(\cdot)\) giving the quantized utilities \(Q^{(t)}_i\) which are then entropy encoded (\textsc{Enc}) before being sent to the CEO (\textsc{Dec}). Based on the received quantized utilities, the CEO updates the quantization function at the users until the desired function has been computed.}
		\label{fig:system-diagram}
	\end{figure}

	We view quantization as a function that maps the finite support set \(\mathcal{X} \subset \mathds{R}\) onto another finite set \(\hat{\mathcal{X}}\), i.e., \(Q: \mathcal{X} \to \hat{\mathcal{X}}\).
	Traditionally, quantization is used as a lossy compression scheme for representing sources with values drawn from a continuous support set.
	In this case, the quantizer is specified by partitioning the support set into intervals that are mapped to representative values.
	Implicit is the assumption that the quantizer is monotonically increasing \(x,y \in \mathcal{X} \; \mathrm{ s.t.\ } x \leq y \implies Q(x) \leq Q(y)\).
	When \(\mathcal{X}\) is finite (as we assume in this work), the quantizer can still be specified in terms of intervals but this representation may not be unique.
	With the assumptions of \(\mathcal{X}\) finite and \(Q\) order preserving, we can alternately specify a \(K\)-level quantizer as a \(K\)-tuple of integers \(\bm{n} = n_1, \ldots, n_K\) that sum to \(L\) with the following interpretation:
	the first \(n_1\) elements of \(\mathcal{X}\) (\(\left\{1, \ldots, n_1\right\}\)) are mapped to \(1\), the next \(n_2\) elements of \(\mathcal{X}\) (\(\left\{n_1 + 1, \ldots, n_2\right\})\) are mapped to \(2\), and so on.
	This representation is unique (in that different integer tuples correspond to different quantization functions) and the set of possible scalar quantizers is isomorphic to the compositions of the integer \(L\) (cf. \defnref{partitions} in \secref{dyn-prog}), which are the sequences of positive integers that sum to \(L\) \cite{Knu2011}.
	For brevity, we define \(n_{1:k} = n_1 + \cdots + n_k\).
	The induced PMF on \(\mathcal{Q}\) is then
	\begin{equation}
		\label{eq:quantizer-pmf}
		p_Q(k) = \sum_{i = n_{1:k-1} + 1}^{n_{1:k}}p_X(i).
	\end{equation}

	We assume that time is slotted into rounds of sufficient length that the CEO can communicate to the users and receive their responses in a single slot.
	We indicate the time slot/round of interaction by \(t\).
	In our analysis, we assume that feedback from the CEO provides all users with the same knowledge as the CEO.

	Suppose that at the beginning of the \(t\)-th round of interaction, the CEO observes that there are \(N^{(t)}\) active users and the support set has size \(L^{(t)}\).
	The CEO will select a quantization function \(Q^{(t)}\) (homogeneous across users) by selecting the number of quantization levels \(K^{(t)}\) and the quantization bin sizes \(\bm{n}^{(t)} = (n^{(t)}_1, \ldots, n^{(t)}_{K^{(t)}})\) and communicate this to the users.
	Let \(Q^{(t)}_i \triangleq Q^{(t)}(X_i)\)	denote the quantization bin in which \(X_i\) lies and let \(\bm{Q}^{(t)} = (Q^{(t)}_i : i = 1, \ldots, N^{(t)})\) denote the length \(N^{(t)}\) tuple of response received by the CEO from the users.
	Define
	\begin{equation}
		N^{(t)}_k = \sum_{i = 1}^{N^{(t)}} \mathds{1}_{Q^{(t)}_i = k}, \quad k = 1, \ldots, K^{(t)}
	\end{equation}
	as the number of sources with state in bin \(k\) and let \(k_t^* = \max \{k : N^{(t)}_k > 0\} = \max_i Q^{(t)}_i\) be the largest of the indices of non-empty quantization bins.
	Based on the responses \(\bm{Q}^{(t)}\) from the active users, the CEO performs the following updates in each round of interaction:
	\begin{subequations}
		\begin{gather}
			N^{(t+1)} = N^{(t)}_{k_t^*}, \qquad L^{(t+1)} = n^{(t)}_{k_t^*}\label{eq:ceo-update}\\
			p^{(t+1)}_X(i) =
			\begin{cases}
				\frac{p^{(t)}_X(i)}{p^{(t)}_Q(k_t^*)} & Q^{(t)}(i) = k^*_t\\
				0 & \textrm{o.w.}
			\end{cases}
		\end{gather}
	\end{subequations}
	The first equation captures the fact that if a user's quantized value is not in the highest reported bin, the user's utility cannot be a maximizing value.
	Only the users with values in the highest reported bin need to continue interacting with the CEO; the remaining users become inactive and do not participate in subsequent rounds.
	The second equation updates the cardinality of the support set; the maximizing value \(X_i\) corresponds to one of the values of \(\mathcal{X}^{(t)}\) that maps to the maximum reported quantization bin.
	The final equation updates the PMF for the remaining range of user values.
	At this point, the CEO is ready to begin the \(t+1\)-th round of interaction.
	Since the initial alphabet of user observations \(L^{(0)}\) is finite and the size of the alphabet decreases in each round \eq{ceo-update}, interaction is guaranteed to terminate after at most \(L^{(0)}\) rounds.
	In the following, we will omit the time superscript when the time instance \(t\) is not relevant to the discussion and/or is clear from context.

	\section{Optimal Solution via Dynamic Programming}
	\label{sec:dyn-prog}
	At time \(t\), the CEO observes the state of the system \(s^{(t)} = \left(N^{(t)}, L^{(t)}, p^{(t)}_X\right)\) and wishes to compute a quantization policy \(\bm{a}^{(t)} \triangleq \left(K^{(t)}, \bm{n}^{(t)}\right) \in \mathcal{A}^{(t)}\) that minimizes the cost of computing the desired function \(f\):
	\begin{dmath}
		\label{eq:cost-dynamic-program}
		C_f\left(\bm{s}^{(t)}\right) = \min_{\bm{a}^{(t)} \in \mathcal{A}^{(t)}} \left[(1 - \lambda)R\left(\bm{a}^{(t)}, \bm{s}^{(t)}\right) + \lambda \tau\left(\bm{s}^{(t)}, \bm{a}^{(t)}\right) + \expected{C_f\left(\bm{S}^{(t+1)}\right) \left\vert \bm{s}^{(t)}\right.}\right].
	\end{dmath}
	The first part of the term inside the minimization consists of a weighting (\(\lambda\)) of the rate and delay incurred by choosing the quantizer given by \(\bm{a}^{(t)}\) when in state \(\bm{s}^{(t)}\).
	The parameter \(\lambda\) is fixed throughout and sets the relative importance of minimizing the rate (\(R(\cdot)\)) versus the delay (\(\tau(\cdot)\)).
	Given \(\bm{s}^{(t)}\) and \(\bm{a}^{(t)}\), the rate and delay are:
	\begin{subequations}
		\begin{align}
			R\left(\bm{a}^{(t)}, \bm{s}^{(t)}\right) &=
			\begin{cases}
				N^{(t)} H\left(p^{(t)}_Q\right) & \bm{s}^{(t)} \not\in \mathcal{S}_f^*\\
				0 & \mathrm{o.w.}
			\end{cases}\\
			\tau\left(\bm{a}^{(t)}, \bm{s}^{(t)}\right)	&=
			\begin{cases}
				1 & \bm{s}^{(t)} \not\in \mathcal{S}_f^*\\
				0 & \mathrm{o.w.}
			\end{cases}
		\end{align}
	\end{subequations}
	where \(p^{(t)}_Q\) is given by \eq{quantizer-pmf} and the set \(\mathcal{S}^*\) represents terminating states (from which \(f\) may be computed).
	In general, the set \(\mathcal{S}^*\) will depend on the particular extremum function being considered.
	If \(\bm{s}^{(t)} \not\in \mathcal{S}^*\), then the CEO will select a quantizer and the rate incurred is the entropy of the induced PMF on the quantized values times the number of active users \(N^{(t)}\); the delay is an additional round of interaction.
	If \(\bm{s}^{(t)} \in \mathcal{S}^*\), the CEO can compute \(f\) and the interaction is over.

	The second part of the expression inside of the minimization consists of the ``cost to go''.
	Depending on the particular state \(\bm{s}^{(t)}\) and particular action \(\bm{a}^{(t)}\) chosen, the system state transitions to state \(\bm{s}^{(t+1)}\) with some probability.
	The ``cost to go'' is an expectation of the optimal cost function taken over all reachable next time step states given the current state \(\bm{s}^{(t+1)}\) and is given by
	\begin{dmath}
		\expected{C_f\left(\bm{S}^{(t+1)}\right) \left\vert \bm{s}^{(t)}\right.} = \sum_{k = 1}^{K^{(t)}}\left[\sum_{i = 1}^{N^{(t)}}\rho(k, i) C_f\left(\left(i, n^{(t)}_k, p^{(t+1)}_X \right)\right)\right]
	\end{dmath}
	where
	\begin{equation}
		\rho(k, i) = \left(p^{(t)}_Q(k)\right)^i\left(\sum_{j=1}^{k-1}p^{(t)}_Q(j)\right)^{N^{(t)} - i}\binom{N^{(t)}}{i}.
	\end{equation}
	The outer summation conditions on the largest reported quantization bin (\(k^*_t = k\)) while the inner summation conditions on there being \(i\) users in the \(k^*_t\)-th bin.
	Given these two outcomes, the state in the next time step is given as \(\bm{s}^{(t+1)} = \left(i, n^{(t)}_k, p^{(t+1)}_X \right)\).

	The next result characterizes the set \(\mathcal{S}^*\) of terminating states for the functions \(\argmax\), \(\max\), or the pair \((\argmax, \max)\).
	\begin{prop}
		\label{prop:argmax-stop}
		When the CEO wishes to determine the \(\argmax\) of the set of users' values, \(\bm{s}^{(t)} \in \mathcal{S}_A^*\) iff \(N^{(t)} = 1\) or \(L^{(t)} = 1\).
	\end{prop}
	\begin{IEEEproof}
		If at some time \(t\) there is only one user still contending for the resource it must be the unique maximizer.
		If the set of possible values consists of a single value, then all remaining users' values equal this value and they are all maximizers.
		In either case, the CEO has losslessly determined the set of \(\argmax\) users.
	\end{IEEEproof}
	\begin{prop}
		\label{prop:max-stop}
		When the CEO wishes to determine either the \(\max\) or the pair \((\argmax, \max)\) of the set of users' values, \(\bm{s}^{(t)} \in \mathcal{S}_M^*\) iff \(L^{(t)} = 1\).
	\end{prop}
	\begin{IEEEproof}
		If the set of possible values consists of a single value, then all remaining users' values equal this value and they are all maximizers.
	\end{IEEEproof}
	The set of terminating states is larger when the CEO wishes to determine the \(\argmax\) because communication can stop when a single user is left, regardless of the set of remaining possible values.
	When determining the \(\max\), the CEO still needs subsequent rounds of communication with the single remaining user to determine its value.
	The following gives the optimal cost \(C_M^{(t)}\) for \(\bm{s}^{(t)} \in \mathcal{S}^*_A \setminus \mathcal{S}^*_M\).
	\begin{prop}
		\label{prop:single-user-cost}
		For \(\bm{s}^{(t)} \in \mathcal{S}^*_A \setminus \mathcal{S}^*_M\), the optimal quantization strategy is for the single remaining user to entropy code the state, and thus complete the communication in one additional round, i.e., \(C^{(t)}_M\left(\bm{s}^{(t)}\right) = (1 - \lambda) H\left(p^{(t)}_X\right) + \lambda\).
	\end{prop}
	\begin{IEEEproof}
		First, we show that for any quantizer \(\bm{n}\) the expected rate is \(H\left(p^{(\delta)}_X\right)\).
		We proceed by induction.
		The base case of \(L^{(t)} = 2\) is immediate.

		For the inductive step, suppose we have a quantizer \(\bm{n} = \left(n_1, \ldots, n_K\right)\); then
		\begin{equation}
			R(\bm{s}^{(t)}) = H(p^{(t)}_Q) + \expected{R(\bm{S}^{(t+1)}) \left\vert\bm{s}^{(t)}\right.}
		\end{equation}
		where
		\begin{equation}
			H(p_Q^{(t)}) = -\sum_{k = 1}^{K^{(t)}} p^{(t)}_Q(k) \log_2 p^{(t)}_Q(k)
		\end{equation}
		and with the inductive assumption it can be shown that
		\begin{dmath}
			\expected{R(\bm{S}^{(t+1)}) \left\vert\bm{s}^{(t)}\right.} = \sum_{k = 1}^{K^{(t)}} p^{(t)}_Q(k) \log_2 p^{(t)}_Q(k) - \sum_{i=1}^{L^{(t)}} p^{(t)}_X(i) \log_2 p^{(t)}_X(i).
		\end{dmath}
		We conclude that \(R(L^{(t)}) = H\left(p^{(t)}_X\right)\), and the cost depends only on the delay.
		A minimum delay of 1 is achieved by \(\bm{n} = (1, \ldots, 1)\).
	\end{IEEEproof}
	The next proposition shows how the cost of computing the \(\argmax\) is related to the cost of computing the \(\max\).
	\begin{prop}
		\label{prop:max-upper-bound}
		The cost for the CEO to compute the \(\max\) exceeds the cost of computing the \(\argmax\), but the difference between the two costs goes to zero as the number of users \(N\) increases.
		We have
		\begin{equation}
			C^{(t)}_A\left(\bm{s}^{(t)}\right) \leq C^{(t)}_M\left(\bm{s}^{(t)}\right) \leq C^{(t)}_A\left(\bm{s}^{(t)}\right) + \bar{\Delta}
		\end{equation}
		where \(\bar{\Delta} \to 0\) as \(N \to \infty\).
	\end{prop}
	\begin{IEEEproof}
		Recall that \(\mathcal{S}^*_M \subset \mathcal{S}^*_A\) (\propref{argmax-stop} \& \propref{max-stop}).
		Therefore, an optimal quantization policy for computing the \(\max\) is a feasible quantization policy for computing the \(\argmax\) and the lower bound is immediate.

		To establish the upper bound, an optimal quantization policy for the \(\argmax\) can be extended into a feasible quantization policy for the \(\max\).
		Starting at time \(t = 0\), the CEO follows the optimal quantization policy for \(\argmax\) until some time \(t = \delta\) such that \(\bm{s}^{(\delta)} \in \mathcal{S}^*_A\).
		If \(\bm{s}^{(\delta)} \in \mathcal{S}^*_M\), then the CEO has determined the \(\max\).
		If \(\bm{s}^{(\delta)} \in \mathcal{S}^*_A \setminus \mathcal{S}^*_M\), then the CEO has determined who the unique maximizer is but not what their value is.
		Since there is a single user left, the minimum cost \(C^{(\delta)}\left(\bm{s}^{(\delta)}\right) = (1 - \lambda) H\left(p^{(\delta)}_X\right) + \lambda\) (cf.\ \propref{single-user-cost}).
		We have
		\begin{equation}
			H\left(p^{(\delta)}_X\right) \leq \log_2 L^{(\delta)} \leq \log_2 \left(L^{(0)} - 1\right)
		\end{equation}
		where the last inequality follows from observing that the size of the support set decreases by at least one at each round and \(\delta \geq 1\).
		As this is a \emph{feasible} quantization policy for the \(\max\) policy we have
		\begin{equation}
			\begin{aligned}
				C^{(t)}_M
				&\leq C^{(t)}_A + \sum_{\bm{s}^{(\delta)} \in \mathcal{S}^*_A \setminus \mathcal{S}^*_M} \prob{\bm{s}^{(\delta)}}\left[(1 - \lambda) H\left(p^{(\delta)}_X\right) + \lambda\right]\\
				&\leq C^{(t)}_A + \prob{\mathcal{E}}\left[(1 - \lambda) \log_2 \left(L^{(0)} - 1\right) + \lambda\right]
			\end{aligned}
		\end{equation}
		where \(\mathcal{E}\) is the event \(\left\{\bm{s}^{(\delta)} \in \mathcal{S}^*_A \setminus \mathcal{S}^*_M\right\}\) and \(\prob{\bm{s}^{(\delta)}}\) is the probability of the quantization policy ending in the state \(\bm{s}^{(\delta)}\).
		We have that \(\mathcal{E} \subseteq \left\{\left|\argmax_i X_i\right| = 1\right\}\).
		Suppose that there are two or more maximizers.
		Then, for any sequence of quantizers, the set of maximizers will respond with the same quantized utility and the only possible terminating state is \(\bm{s}^{(\delta)} \in \mathcal{S}^*_M\).

		To show that \(\bar{\Delta} \to 0\) as \(N \to \infty\) it suffices to show that
		\begin{equation}
			\lim_{N \to \infty} \prob{\left|\argmax_i X_i\right| = 1} = 0.
		\end{equation}
		Let \(A\) denote the event that \(\left|\argmax_i X_i\right| = 1\) and \(B_k\) denote the event \(\max_i X_i = k\).
		By the law of total probability we have
		\begin{equation}
			\prob{A} = \sum_{k = 1}^{L} \prob{A \cap B_k} = \sum_{k = 1}^{L} N F_X(k - 1)^{N - 1}p_X(k).
		\end{equation}
		Since
		\begin{equation}
			\lim_{N \to \infty} N F^{N-1}_X(k-1) p_X(k) = 0 \quad \forall \; k = 1, \ldots, L,
		\end{equation}
		it follows that \(\prob{A} \to 0\) as \(N \to \infty\)
	\end{IEEEproof}

	To trace out the rate-delay trade-offs of interactive scalar quantization, the dynamic program of \eq{cost-dynamic-program} is solved for multiple values of \(\lambda\) and the corresponding rate-delay values are plotted parametrically.
	This is done in \secref{results} for representative source distributions.
	The following definitions of two related combinatorial objects are needed before discussing an assumption concerning the search space \(\mathcal{A}^{(t)}\) in \eq{cost-dynamic-program}.
	\begin{defn}[Compositions \& partitions of an integer \cite{Knu2011}]
		\label{defn:partitions}
		The \emph{compositions of an integer} \(L\) are the sequences of positive integers that sum to \(L\).
		The \emph{partitions of an integer} are the ways to write it as a sum of positive integers, disregarding order.
	\end{defn}
	For example, consider the integer 3: there are 4 compositions (namely \(\{(3), (2, 1), (1, 2), (1, 1, 1)\}\)) while there are only 3 partitions (\(\{(3), (2, 1), (1, 1, 1)\}\)).
	\begin{assmp}
		\label{assmp:partitions}
		We take the set of quantizers \(\mathcal{A}^{(t)}\) in solving \eq{cost-dynamic-program} to be the set of all partitions of the integer \(L\) instead of the set of all compositions.
	\end{assmp}
	\begin{rem}
		As a justification for the above assumption, recall that in \secref{problem-model} we showed that set of quantizers is isomorphic to the set of compositions of the integer \(L\).
		Using basic combinatorial arguments, it can be shown that the number of compositions \(c(L)\) for a given integer \(L\) (and thus the number of possible quantizers) is
		\begin{equation}
			\label{eq:composition-count}
			c(L) = 2^{L-1}.
		\end{equation}
		G.\ H.\ Hardy and S. Ramanujan \cite{Knu2011} obtained the following expression for the asymptotic behavior for the number of partitions \(\varphi(L)\)
		\begin{equation}
			\label{eq:partition-count}
			\varphi(L) \sim \frac{1}{4L\sqrt{3}}\exp\left(\pi\sqrt{\frac{2L}{3}}\right) \quad n \to \infty.
		\end{equation}
		For computational tractability, we take the set of quantizers to be the set of all partitions instead of compositions.
		Our justification for doing so is shown in \fig{composition-partition-comparison}.
		For a small initial number of users \(N\) and initial support set size \(L = 16\), we computed the rate and delay cost components for every composition and every partition assuming subsequent rounds are solved optimally.
		The Pareto optimal boundary for compositions (markers) matches the boundary for partitions (no markers) for these parameters.
		\begin{figure*}
			\centering
			\includegraphics{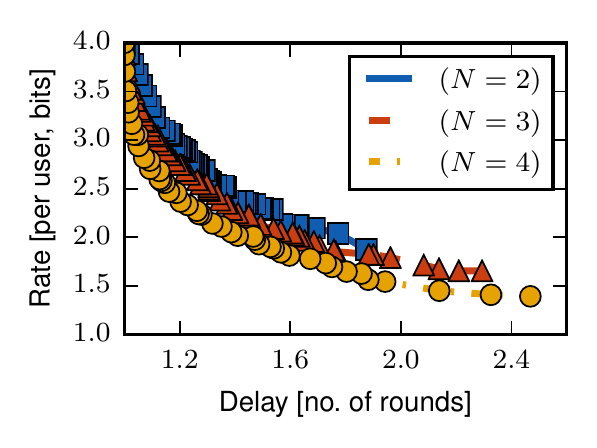}
			\caption{Comparison of compositions and partitions. Shown is the rate-delay trade-off obtained by optimizing over the set of compositions (\emph{markers}) and partitions (\emph{no markers}) for uniform sources with support set size \(L = 16\).}
			\label{fig:composition-partition-comparison}
		\end{figure*}

		For a given source distribution, as the number of users increases, the distribution of the maximum becomes more and more ``peaked'' about the largest possible value; this is made more rigorous in \lemref{almost-sure}.
		Since the CEO is seeking to identify either the \(\argmax\) or \(\max\) (or both), we expect that quantizers that more finely quantize the larger values of the support set will outperform those quantizers that do not.
	\end{rem}
	\begin{lem}
		\label{lem:almost-sure}
		Suppose \(X_i \sim p_X(x)\) where \(x\) takes values in the finite alphabet \(\mathcal{X} = \{x_1, \ldots, x_L\}\), then the sequence \(Y_n = \max_{1 \leq i \leq n} X_i\) converges almost surely towards \(x_L\).
	\end{lem}
	\begin{IEEEproof}
		It is easily established that \(P_n(\epsilon) = \mathds{P}\left(\left\vert Y_n - x_L\right\vert > \epsilon\right) \leq \mathds{P}\left(Y_n \leq x_{L-1}\right)\) for all \(\epsilon > 0\) from which it follows
		\begin{equation}
			\sum_n P_n(\epsilon) \leq \frac{1 - p_X(x_L)}{p_X(x_L)} < \infty
		\end{equation}
		and therefore \(Y_n\) converges almost surely towards \(x_L\) (cf. Theorem 7.2.4 in \cite{GriSti1992}).
	\end{IEEEproof}

	In formulating the cost function of \eq{cost-dynamic-program}, only the rate on the uplink (i.e., users to CEO) was considered.
	The reason for focusing exclusively on the uplink costs are concerns of asymmetric power constraints when applying this framework to cellular systems.
	In this context, the users are the battery-constrained mobile stations while the CEO is the basestation.
	This motivates the next assumption concerning \eq{cost-dynamic-program}.
	\begin{assmp}
		We assume the cost of communication from users to the CEO is more expensive than communication from the CEO to the users.
		Therefore we omit the cost of dissemination on the downlink from the CEO to the users in our analysis.
	\end{assmp}
	\begin{rem}
		As a justification for the above assumption, note that after collecting all the responses from the users, the CEO observes the next state \(\bm{S}^{(t+1)}\) and needs to convey this back to the users.
		Assuming the CEO broadcasts feedback to the users, the rate of feedback \(R_{\text{CEO}}\) bounded from above by
		\begin{equation}
			R_{\text{CEO}} \leq H\left(\bm{S}^{(t)} | \bm{s}^{(t)}\right).
		\end{equation}
		Observe that this entropy over-estimates the rate required: at the end of interaction, not only does the CEO know the value of the function, but so do all the users.
		Define
		\begin{equation}
			\hat{\bm{S}}^{(t+1)} =
			\begin{cases}
				\bm{S}^{(t+1)} & \bm{S}^{(t+1)} \not\in \mathcal{S}_f^*\\
				0 & \bm{S}^{(t+1)} \not\in \mathcal{S}_f^*\\
			\end{cases}
		\end{equation}
		which condenses all of the terminating states for the function \(f\) into one.
		This allows the CEO to signal the end of interaction, without the extra rate to convey the function result back to users.
		To compute an upper bound on the optimal cost including the rate of feedback from the CEO, the objective function of \eq{cost-dynamic-program} is changed to
		\begin{dmath}
			\label{eq:cost-dynamic-program-feedback}
			\hat{C}_f\left(\bm{s}^{(t)}\right) = \min_{\bm{a}^{(t)} \in \mathcal{A}^{(t)}} \left[(1 - \lambda)\left(R\left(\bm{a}^{(t)}, \bm{s}^{(t)}\right) + H\left(\hat{\bm{S}}^{(t)} | \bm{s}^{(t)}\right)\right) + \lambda \tau\left(\bm{s}^{(t)}, \bm{a}^{(t)}\right) + \expected{C_f\left(\bm{S}^{(t+1)}\right)}\right].
		\end{dmath}

		\fig{ceo-feedback-cost} shows the impact on the rate-delay trade-offs of the optimal scalar quantization policy.
		\begin{figure*}
			\includegraphics{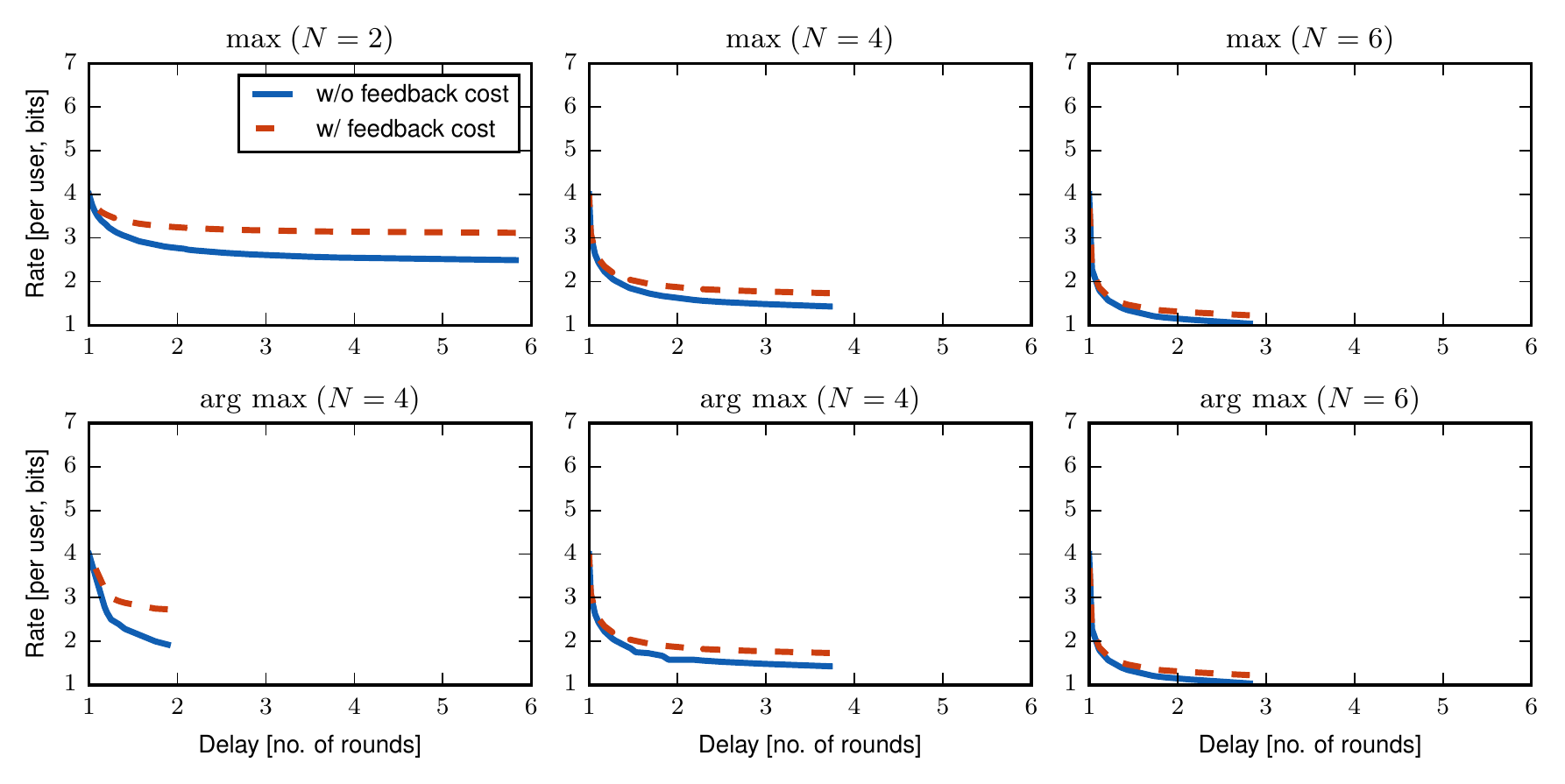}
			\caption{Comparison of rate-delay trade-offs with and without including the cost of feedback from the CEO to the users. The source distribution was uniform with support set size \(L = 16\).}
			\label{fig:ceo-feedback-cost}
		\end{figure*}
		The effect is to increase the required rate for a given delay (or increase the required delay for a given rate), with the effect being most pronounced when the number of users \(N\) is small.
		These results are for the case where the cost of communication from the users to the CEO equals the cost of communication from the CEO to the users.
		A more nuanced approach would modify \eq{cost-dynamic-program-feedback} to include a weighting factor to adjust the relative importance of the rate from the users to the CEO versus the rate from the CEO to the users.
		Since the impact of including the rate of feedback is already marginal, this would only further reduce the differences in the rate-delay trade-off curves.
	\end{rem}

	\section{Analysis of Suboptimal Schemes}
	\label{sec:simple-schemes}
	The dynamic programming formulation of the previous section is amenable to computing the minimum cost, and therefore the rate-delay trade-off, of interactive scalar quantization as an achievable scheme.
	A drawback with computing the solution to the dynamic program is that it does not provide insight into how the minimum cost scales in the number of users \(N\) and/or support set size \(L\).
	Additionally, the computation provides little insight about the structure of the optimal quantizers.
	In this section, we consider two simple quantization strategies and derive expressions for the associated rate and delay.
	We then generalize to a family of strategies and prove their near-optimality; the significant reduction in the size of the search space results in faster computation of \eq{cost-dynamic-program} with only a small penalty.

	In this section, we assume the users' utilities have a uniform distribution.
	This assumption is motivated by the results of \secref{results-distributions} and the analytical tractability of the resulting rate and delay expressions.

	\subsection{Binary search}
	We first consider a quantization strategy for computing the \(\argmax\) inspired by binary search.
	At each round, the remaining support set is divided in half and the users indicate whether their values lie in the lower or upper half.
	This process is repeated until either a single user remains or the support set has been reduced to one.

	We assume \(L\) is a power of two to repeatedly divide in half.
	The rate for this scheme is given by
	\begin{equation}
		\label{eq:binary-search-rate-recursion}
		R_b(N, 2L) = N + \frac{R_b(N, L) + \sum_{i = 2}^{N} \binom{N}{i}R_b(i, L)}{2^N}
	\end{equation}
	with base cases \(R_b(1, \cdot) = R_b(\cdot, 1) = 0\).
	Since the initial support set size (\(2L\)) is being halved, each user replies with a single bit for a total of \(N\) bits.
	Depending on the users' values, all \(N\) users could be in the lower half (\(L\)) which happens with probability \(2^{-N}\); or \(i\) users could be in the upper half which happens with probability \(\binom{N}{i}2^{-N}\).
	Following a similar line of reasoning, delay is given by
	\begin{equation}
		\label{eq:binary-search-delay-recursion}
		\tau_b(N, 2L) =  1 + \frac{\tau_b(N, L) + \sum_{i = 2}^{N} \binom{N}{i}\tau_b(i, L)}{2^N},
	\end{equation}
	with base cases \(\tau_b(1, \cdot) = \tau_b(\cdot, 1) = 0\).
	\begin{prop}
		\label{prop:binary-search}
		If \(L\) is a power of two, the expected rate of computing the \(\argmax\) with binary search is
		\begin{equation}
			\label{eq:binary-search-rate}
			R_b(N, L) = 2 N \left(1 - \frac{1}{L}\right)\\
		\end{equation}
		and the expected delay is bounded by
		\begin{equation}
			\label{eq:binary-search-delay}
			\tau_b(N, L) \leq \min\{ \log_2 N + 1, \log_2 L \}
		\end{equation}
		for \(N \geq 2\) and \(L \geq 2\).
		It follows that
		\begin{equation}
			\lim_{L \to \infty} \frac{R_b(N, L)}{N} = 2, \qquad \lim_{L \to \infty} \tau_b(N, L) \leq \log_2 N + 1.
		\end{equation}
	\end{prop}
	\begin{IEEEproof}
		We prove the expression for expected rate by induction.
		The base case of \(N = 2\) and \(L = 2\) is immediate.
		For the inductive step, we have
		\begin{equation}
			\begin{split}
				&R_b(N, 2L)\\
				&\stackrel{(a)}{=} N + 2^{-N} \left[2 N \left(1 - \frac{1}{L}\right) + \sum_{i = 2}^{N} \binom{N}{i}\left(2 i \left(1 - \frac{1}{L}\right)\right)\right]\\
				&\stackrel{(b)}{=} N + 2^{-(N-1)} \left(1 - \frac{1}{L}\right)2^{(N - 1)} N
			\end{split}
		\end{equation}
		where (a) follows by the inductive assumption and \eq{binary-search-rate-recursion} \& \eq{binary-search-rate}, and (b) follows from a standard identity.

		To prove the upper bound for delay, we first consider the case of \(L\) fixed and show that
		\begin{equation}
			\tau_b(N, L) \leq \log_2 L \quad \forall \; N.
		\end{equation}
		We proceed by induction in \(L\); the base case of \(L = 1\) is immediate.
		For the inductive step, we have
		\begin{equation}
			\begin{split}
				&\tau_b(N, 2L)\\
				&\stackrel{(a)}{\leq} 1 + 2^{-N}\left[\log_2 L + \sum_{i = 2}^{N} \binom{N}{i} \log_2 L\right]\\
				&\stackrel{(b)}{=} 1 + \left(1 - N2^{-N}\right) \log_2 L \leq \log_2 \left(2L\right)
			\end{split}
		\end{equation}
		where (a) follows by the inductive assumption and \eq{binary-search-delay-recursion} \& \eq{binary-search-delay}, and (b) follows from a standard identity.

		We now consider the case of \(N\) fixed and show that
		\begin{equation}
			\tau_b(N, L) \leq \log_2 N + 1 \quad \forall \; L.
		\end{equation}
		We proceed by induction in \(N\); the base case of \(N = 1\) is immediate.
		For the inductive step, let
		\begin{equation}
			g(i) \triangleq \log_2 (2i) = \log_2 i + 1.
		\end{equation}
		Observe that the recurrence for \(\tau_b(\cdot, \cdot)\) can be written as
		\begin{equation}
			\tau_b(N, 2L) = 1 + \expected{\tau_b(Z, L)}
		\end{equation}
		where \(Z\) is a random variable with PMF given as
		\begin{equation}
			\prob{Z = i} =
			\begin{cases}
				\binom{N}{i}2^{-N} & i \in \{1, \ldots, N - 1\}\\
				2^{-(N - 1)} & i = N
			\end{cases}
		\end{equation}
		and expected value
		\begin{equation}
			\expected{Z} = N\left(\frac{1}{2} + 2^{-N}\right).
		\end{equation}
		Applying the law of total expectation we have
		\begin{equation}
			\tau_b(N, 2L) = 1 + \expected{\tau_b(Z, L) \middle| Z > 1}\prob{Z > 1}
		\end{equation}
		which follows from \(\tau_b(1, L) = 0\).
		Substituting the upper bound gives
		\begin{equation}
			\tau_b(N, 2L) \leq 1 + \expected{g(Z) \middle| Z > 1}\left(1 - N2^{-N}\right).
		\end{equation}
		We have
		\begin{equation}
			\begin{split}
				&\expected{g(Z) \middle| Z > 1} \stackrel{(a)}{\leq} g\left(\expected{Z \middle| Z > 1}\right)\\
				&\stackrel{(b)}{=} g\left(\frac{N}{2\left(1- N 2^{-N}\right)}\right) = \log_2\left(\frac{N}{1-N 2^{-N}}\right)
			\end{split}
		\end{equation}
		where (a) follows by Jensen's inequality and (b) follows from
		\begin{equation}
			\begin{split}
				&\expected{Z \middle| Z > 1} = \frac{\expected{Z} - \expected{Z \middle| Z = 1}\prob{Z = 1}}{\prob{Z > 1}}\\
				&= \frac{N\left(\frac{1}{2} + 2^{-N}\right) - N 2^{-N}}{1 - N 2^{-N}} = \frac{N}{2\left(1- N 2^{-N}\right)}.
			\end{split}
		\end{equation}
		It follows that
		\begin{equation}
			\expected{g(Z) \middle| Z > 1}\left(1 - N2^{-N}\right) \leq \log_2 N
		\end{equation}
		and we conclude \(\tau_b(N, 2L) \leq 1 + \log_2 N\).
	\end{IEEEproof}
	Binary search is an attractive quantization strategy because it requires at most 2 bits per user on average to compute the \(\argmax\) and the delay remains bounded (by the number of users) as the support set size grows.
	With the support set being halved at each round of interaction, the support set size does not equal one until after \(\log_2 L\) rounds of interaction.
	Using binary search as a quantization strategy for computing the \(\max\) has a delay of exactly \(\log_2 L\) rounds (c.f.\ \propref{max-stop}).

	\subsection{Max search}
\begin{figure}
		\centering
		\includegraphics{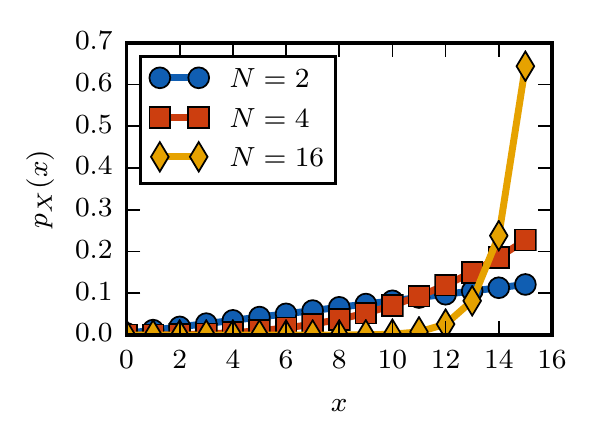}
		\caption{Probability mass function for \(\max_i X_i\) when \(X_i\)'s are i.i.d.\ uniform (\(L = 16\)) as a function of the number of users.
		As \(N\) increases, the probability becomes more concentrated around the larger values.}
		\label{fig:pmf-concentration}
	\end{figure}

	As described at the end of the previous subsection, binary search has a constant delay of \(\log_2 L\) rounds when the CEO wishes to compute the \(\max\).
	We propose the following search strategy, which we refer to as \emph{max search}:
	at each round, the users indicate to the CEO whether their source observation is the largest possible value in the current support.
	If at least one user replies in the affirmative, the interaction stops and the \(\max\) has been found.
	Only if none of the users have the largest value does the interaction continue.
	This strategy is motivated by the following two observations:
	\begin{inparaenum}
		\item for a fixed support set size, as the number of users increases, the PMF of the \(\max\) becomes more and more peaked about the larger support set values (cf.\ \lemref{almost-sure} \& \fig{pmf-concentration}), and;
		\item at every round of interaction, there is a quantization bin of size one and therefore a non-zero probability of the interaction ending in the current round.
	\end{inparaenum}

	The rate for max search can be written recursively as
	\begin{equation}
		\label{eq:max-search-rate-recursion}
		R_m(N, L) = N h(p_L) + (1 - p_L)^N R_m(N, L - 1)
	\end{equation}
	where \(p_L = 1/L\); the delay can be written recursively as
	\begin{equation}
		\label{eq:max-search-delay-recursion}
		\tau_m(N, L) = 1 + (1 - p_L)^N \tau_m(N, L - 1).
	\end{equation}

	\begin{prop}
		\label{prop:max-search}
		We have
		\begin{subequations}
			\begin{align}
				R_m(N, L)
				&= N \sum_{i=2}^L \left(\frac{i}{L}\right)^N h(1/i) \label{eq:max-search-rate}\\
				\tau_m(N, L)
				&= \sum_{i = 2}^{L} \left(\frac{i}{L}\right)^N \label{eq:max-search-delay}.
			\end{align}
		\end{subequations}
		It follows that
		\begin{equation}
			\lim_{N \to \infty} \frac{R_m(N, L)}{N} = h\left(\frac{1}{L}\right), \quad \lim_{N \to \infty} \tau_m(N, L) = 1.
		\end{equation}
	\end{prop}
	\begin{IEEEproof}
		By induction on \(L\).
		The base case of \(L = 2\) is immediate.
		For the inductive step we have
		\begin{equation}
			\begin{split}
				&R_m(N, L)\\
				&= N h(1/L) + \left(\frac{L-1}{L}\right)^N N \sum_{i=1}^{L - 1} \left(\frac{i}{L-1}\right)^N h(1/i)\\
			\end{split}
		\end{equation}
		which follows from the inductive assumption and \eq{max-search-rate-recursion} \& \eq{max-search-rate}.
		The proof for \(\tau_m\) follows the same arguments.
	\end{IEEEproof}
	This scheme has very low rate and delay as the number of users \(N\) gets larger.
	Using this scheme, however, the CEO does not have the ability to select a desired rate-delay trade-off.
	In the next subsection, we extend this simple search strategy to a family of search strategies that give the CEO the ability to operate at a desired rate-delay trade-off.
	We conclude by noting that even though max search is designed to enable the CEO to interactively compute the \(\max\), \propref{max-upper-bound} establish it as a quantization strategy for computing the \(\argmax\) as well.

	\subsection{Extended max search}
	We extend the quantization strategy of the previous section into a family of quantization strategies.
	The previous strategy worked by asking the users to indicate whether or not their state \(X_i = \max \mathcal X^{(t)}\) at each iteration \(t\), terminating when at least one user replied in the affirmative.
	We extend this strategy by asking the users to indicate which of the \(K - 1\) largest values of \(\mathcal{X}^{(t)}\) they have or indicating their state is not one of these values.
	For example: consider \(L = 5\) and \(K = 4\).
	The quantizer for this parameter set would be \(\bm{n} = (2, 1, 1, 1)\).
	Like the quantization strategy of the previous section, this family of strategies has the property that iteration continues if and only if all \(N\) users' states are in the first bin.
	Unlike the previous strategy, we are not able to write a closed-form expression for the rate and delay components when selecting quantizers from this family of quantizers.
	However, we are able to prove several non-trivial and important properties of this family.

	We begin by giving a formal description of the family.
	For notational compactness, let \(\bm{1}_k\) be the \(k\)-tuple of all ones and denote tuple concatenation as \(\oplus\).
	Suppose we are in a state \(\bm{s} = (N, L)\).
	Define
	\begin{equation}
		\label{eq:max-search-family}
		\mathcal{L}(L) = \left\{(L - k + 1) \oplus \bm{1}_{k-1} : k = 2, \ldots, L\right\}.
	\end{equation}
	For example: \(\mathcal{L}(4) = \left\{(3, 1), (2, 1, 1), (1, 1, 1, 1)\right\}\).
	Our first result shows that for any quantizer \(\bm{n} \in \mathcal{L}(L)\), permutation of the bin sizes results in a quantizer with a higher cost.
	Recall Assumption~\ref{assmp:partitions} of \secref{dyn-prog} where we took the search space of quantizers to be the set of partitions instead of compositions.
	Our justifications for this assumption were concerns of computational complexity (cf. \eq{composition-count} \& \eq{partition-count}) and the observation that the optimal solution was still found when searching over partitions for small problem instances (\fig{composition-partition-comparison}, right).
	Since permutation of an integer partition gives (in general) a composition, the next result is a further justification of Assumption~\ref{assmp:partitions}.
	\begin{prop}
		For computing the \(\max\) of the users' values, the \(K\)-level quantizer \(\bm{n} = \left(L - K + 1\right) \oplus \bm{1}_{K-1}\) has lower cost than any other quantizer obtained by permutation.
	\end{prop}
	\begin{IEEEproof}
		Let \(\bm{n}_m\) be the quantizer obtained from \(\bm{n}\) by shifting the bin of size \(L - K + 1\) \(m\) locations to the right.
		For example \(\bm{n}_3 = \bm{1}_3 \oplus (L - K + 1) \oplus \bm{1}_{K-4}\).
		The stage cost of the quantizer
		\begin{equation}
			(1 - \lambda) R\left(\bm{a}^{(t)}, \bm{s}^{(t)}\right) + \lambda \tau\left(\bm{a}^{(t)}, \bm{s}^{(t)}\right)
		\end{equation}
		is invariant to permutation.
		The expected cost to go of the quantizer \(\bm{n}\) is
		\begin{equation}
			\left(\frac{L - K + 1}{L}\right)^{N} C_M^{(t+1)}\left(N, L - K + 1\right)
		\end{equation}
		and the expected cost to go of the quantizer
		\(\bm{n}_m\) is
		\begin{equation}
			\begin{gathered}
				\sum_{i = 1}^{N} \rho(i) C_M^{(t+1)}\left(i, L - K + 1\right)\\
				\rho(i) = \binom{N}{i}\left(\frac{m}{L}\right)^{N - i}\left(\frac{L - K + 1}{L}\right)^{i}
			\end{gathered}
		\end{equation}
		Taking the difference we see
		\begin{equation}
			\sum_{i = 1}^{N - 1}\rho(i) C_M^{(t+1)}\left(i, L - K + 1\right) \geq 0
		\end{equation}
		and conclude that \(\bm{n}\) has better cost than \(\bm{n}_m\).
		As the choice of \(m\) was arbitrary, the result holds for any permutation.
	\end{IEEEproof}

	An attractive property of \(\mathcal{L}(L)\) is that \(\left\vert\mathcal{L}(L)\right\vert = L - 1\) where as the number of all quantizers is exponential in \(L\).
	If we could show that \(\mathcal{L}(L)\) was \emph{sufficient} for solving \eq{cost-dynamic-program} (instead of the set of all quantizers) optimally, it would represent a significant reduction in computational complexity.
	As a first step, we show for a given quantizer \(\bm{n} \not\in \mathcal{L}(L)\) how to select a quantizer \(\bm{n}_r \in \mathcal{L}(L)\) that asymptotically in \(N\) has performance no worse than \(\bm{n}\).

	Consider a \(K\)-bin quantizer \(\bm{n} = (n_1, \ldots, n_K)\); if \(n_1 = L - K + 1\), then \(\bm{n} \in \mathcal{L}(L)\).
	Otherwise \(n_1 < L - K + 1\) and \(\bm{n} \not\in \mathcal{L}(L)\).
	Let \(\bm{n}_r = (L - K + 1) \oplus \bm{1}_{K-1} \in \mathcal{L}(L)\) be the quantizer with the same number of bins as \(\bm{n}\).
	For example: if \(\bm{n} = (2, 2, 1)\) then \(\bm{n}_r = (3, 1, 1)\).

	For notational compactness in the rest of the section, denote the current state as \(\bm{s} = (N, L)\) (which is fixed and known) and the state in the next iterations as \(\bm{s}' = (N', L')\) (which are discrete random variables whose PMF depends on the select quantizer).
	The cost when using \(\bm{n}\) is
	\begin{equation}
		C_{\bm{n}}(N, L) = (1 - \lambda) N H(\bm{n}) + \lambda + \mathds{E}_{\bm{n}}\left[C(N', L')\right]
	\end{equation}
	where
	\begin{subequations}
		\label{eq:n-cost-to-go}
		\begin{equation}
			\mathds{E}_{\bm{n}}\left[C(N', L')\right] = \sum_{j = 1}^k\sum_{i = 1}^N\mathds{P}_{\bm{n}}\left[N' = i, L' = n_j\right]C\left(i, n_j\right),
		\end{equation}
		and
		\begin{equation}
			\begin{split}
				&\mathds{P}_{\bm{n}}\left[N' = i, L' = n_j\right] =\\
				&\left[\left(\frac{n_{1:j}}{L}\right)^N - \left(\frac{n_{1:j-1}}{L}\right)^N\right]\frac{\binom{N}{i} \rho_j^i\left(1 - \rho_j\right)^{N - i}}{1 - (1 - \rho_j)^N},
			\end{split}
		\end{equation}
		and \(\rho_j = \frac{n_j}{n_{1:j}}\).
	\end{subequations}
	The cost when using \(\bm{n}_r\) is
	\begin{equation}
		C_{\bm{n}_r}(N, L) = (1 - \lambda) N H(\bm{n}_r) + \lambda + \mathds{E}_{\bm{n}_r}\left[C(N', L')\right]
	\end{equation}
	where
	\begin{equation}
		\label{eq:n_r-cost-to-go}
		\mathds{E}_{\bm{n}_r}\left[C(N', L')\right] = \left(\frac{L - K + 1}{L}\right)^N C(N, L - K + 1).
	\end{equation}
	Taking the difference we have
	\begin{equation}
		\label{eq:n-n_r-delta}
		\begin{split}
			&\Delta_{\bm{n}, \bm{n}_r} = C_{\bm{n}}(N, L) - C_{\bm{n}_r}(N, L)\\
			&= (1 - \lambda) N \left(H(\bm{n}) - H(\bm{n}_r)\right)\\
			&+ \mathds{E}_{\bm{n}}\left[C(N', L')\right] - \mathds{E}_{\bm{n}_r}\left[C(N', L')\right].
		\end{split}
	\end{equation}
	The difference in the stage costs of quantizer \(\bm{n}\) and \(\bm{n}_r\) is expressed in terms of the difference in entropies of the induced probability mass functions \(H(\bm{n}) - H(\bm{n}_r) \geq 0\) where the inequality follows from the fact that \(\bm{q}\) is majorized by \(\bm{b}\) and \(x \log x\) is convex \cite{MarOlkArn2011}.

\begin{figure}
		\centering
		\includegraphics{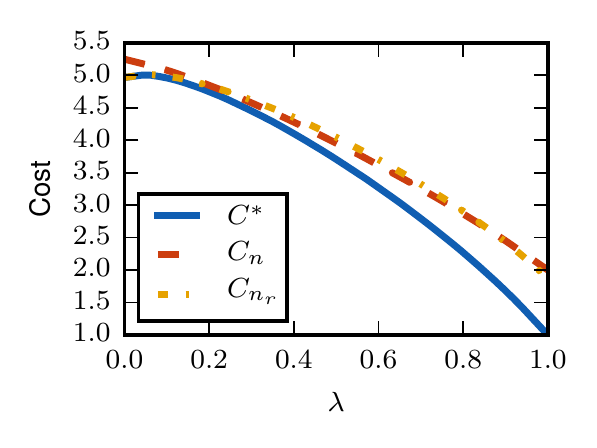}
		\caption{Comparison of the cost for \(\bm{n} = (11, 5)\) and \(\bm{n}_r = (15, 1)\) for the state \(\bm{s} = (N = 2, L = 16)\). \(\bm{n}_r\) does not outperform \(\bm{n}\) in terms of cost for all \(\lambda\).}
		\label{fig:quantizer-row-counter-example}
	\end{figure}

	Unfortunately, the difference in the expected cost to go is not always positive.
	In fact, the difference in the expected cost to go can be negative enough to offset the positive difference in the quantizer rates.
	In \fig{quantizer-row-counter-example} we consider the case of \(N = 2\) and \(L = 16\) and compare the costs of the non-\(\mathcal{L}(16)\) quantizer \(\bm{n} = (11, 5)\) to the \(\bm{n}_r \in \mathcal{L}(16)\) quantizer \((15, 1)\).
	For certain values of \(\lambda\), \(C_{\bm{n}}(N, L)\) is less than \(C_{\bm{n}_r}(N, L)\).
	This does \emph{not} disprove the sufficiency of \(\mathcal{L}(16)\) in solving \eq{cost-dynamic-program} optimally; the range of \(\lambda\) for which \(C_{\bm{n}}(N, L) < C_{\bm{n}_r}(N, L)\) does not coincide with the range of \(\lambda\) for which \(C(N, L) = C_{\bm{n}_r}(N, L)\).
	The above derivation allows us to prove that \(\mathcal{L}(L)\) is asymptotically sufficient for solving \eq{cost-dynamic-program} optimally.
	\begin{prop}
		\label{prop:asymptotically-optimal}
		\(\mathcal{L}(L)\) is asymptotically sufficient for minimizing the cost associated with interactively computing the \(\max\).
		The set \(\mathcal{L}(L)\) \emph{cannot} be made smaller without losing this property.
		For a given value of \(L\) and \(\bm{n} \not\in \mathcal{L}(L)\)
		\begin{equation}
			\lim_{N \to \infty} \Delta_{\bm{n}, \bm{n}_r} \geq 0
		\end{equation}
		where \(\bm{n}_r \in \mathcal{L}(L)\) has the same number of bins as \(\bm{n}\).
	\end{prop}
	\begin{IEEEproof}
		The proof proceeds by induction with an inductive assumption that only quantizers from \(\mathcal{L}(L^{(t)})\) for \(t \geq 1\) are used for subsequent rounds.

		For the state \((N, L)\), \(\Delta_{\bm{n}, \bm{n}_r}\) is asymptotically (in \(N\)) non-negative.
		By assumption \(\bm{n} \not\in \mathcal{L}(L)\), therefore \(L - K + 1 > n_1 \geq n_2 \geq \cdots \geq n_K \geq 1\).
		This, in turn, implies that \(\hat{\bm{n}}_1 = (n_1) \oplus \bm{1}_{L-K+1-n_1} \in \mathcal{L}(L - K + 1)\) is a valid quantizer for the state \(\bm{s}' = (N^{(t)}, L^{(t)} - K + 1)\) and
		\begin{equation}
			\label{eq:bound-1}
			\begin{split}
				&C_M\left(\bm{s}'\right) \leq (1 - \lambda) N^{(t+1)} H(\hat{\bm{n}}_1) + \lambda\\
				&+ \left(\frac{n_1}{L^{(t)} - K + 1}\right)^n C_M(N^{(t)}, n_1).
			\end{split}
		\end{equation}
		Next, there exists some \(k^*\) such that \(n_2 \geq \cdots \geq n_{k^*} > 1\); if not then \(\bm{n} \in \mathcal{L}(L)\).
		We have that
		\begin{equation}
			\label{eq:bound-2}
			C_M(i, n_j)\geq \lambda \quad j \leq k^* \quad C_M(i, n_j) = 0 \quad j > k^*.
		\end{equation}
		Finally, for notational compactness let \(\nu = \frac{L - K + 1}{L} < 1\).

		We then have
		\begin{equation}
			\label{eq:delta-bound}
			\begin{aligned}
				\Delta_{\bm{n}, \bm{n}_r}
				&\stackrel{(a)}{\geq} (1 - \lambda) N \left(H(\bm{n}) - H(\bm{n}_r) - \nu^N H(\hat{\bm{n}}_1)\right)\\
				&+\sum_{j = 1}^{k^*}\sum_{i = 1}^N\mathds{P}_{\bm{n}}\left[N^{(t+1)} = i, L^{(t+1)} = n_j\right]C\left(i, n_j\right)\\
				&- \nu^N\left(\left(\frac{q_1}{L - K + 1}\right)^N C(N, n_1) + \lambda \right)\\
				&\stackrel{(b)}{=} (1 - \lambda) N \left(H(\bm{n}) - H(\bm{n}_r) - \nu^N H(\hat{\bm{n}}_1)\right) - \nu^N \lambda\\
				&+\sum_{j = 2}^{k^*}\sum_{i = 1}^N\mathds{P}_{\bm{n}}\left[N^{(t+1)} = i, L^{(t+1)} = n_j\right]C\left(i, n_j\right)\\
				&\stackrel{(c)}{\geq} (1 - \lambda) N \left(H(\bm{n}) - H(\bm{n}_r) - \nu^n H(\hat{\bm{n}}_1)\right)\\
				&+ \sum_{j = 2}^{k^*}\sum_{i = 1}^N\mathds{P}_{\bm{n}}\left[N^{(t+1)} = i, L^{(t+1)} = n_j\right]\lambda - \nu^N \lambda\\
				&= (1 - \lambda) N \left(H(\bm{n}) - H(\bm{n}_r) - \nu^N H(\hat{\bm{n}}_1)\right)\\
				&+ \lambda \left(\frac{n_{1:k^*}}{L}\right)^N \left[1 - \left(\frac{n_1}{n_{1:k^*}}\right)^N - \left(\frac{L - K +1}{n_{1:k^*}}\right)^N\right]
			\end{aligned}
		\end{equation}
		where
		\begin{enumerate}[(a)]
			\item follows from \eq{n-cost-to-go}, \eq{n_r-cost-to-go}, \eq{n-n_r-delta}, \eq{bound-1}, and \eq{bound-2}
			\item follows from
			\begin{equation}
				\mathds{P}_{\bm{n}}\left[N^{(t+1)} = i, L^{(t+1)} = q_1\right] =
				\begin{cases}
					0, & i < n\\
					\left(\frac{q_1}{L}\right)^n, & i = n
				\end{cases}
			\end{equation}
			and
			\item follows from \eq{bound-2}
		\end{enumerate}
		\(n_{1:k^*} = L - k + k^* > K - k + 1\) and therefore the right hand side of \eq{delta-bound} has a non-negative limit because
		\begin{equation}
			\lim_{N\to\infty} \left(\frac{n_{1:k^*}}{L}\right)^N \left[1 - \left(\frac{n_1}{n_{1:k^*}}\right)^N - \left(\frac{L - K +1}{n_{1:k^*}}\right)^N\right] = 0.
		\end{equation}

		The set \(\mathcal{L}(L)\) contains one and only one quantizer for each possible bin size.
		If this set were smaller, then for a given \(\bm{n}\) the quantizer \(\bm{n}_r\) (which has the same number of bins as \(\bm{n}\)) may not be in \(\mathcal{L}'(L) \subset \mathcal{L}(L)\).
	\end{IEEEproof}

	In general, there exists values of \(N\) and \(L\) for which \(\mathcal{L}(L)\) is not sufficient.
	Let \(\mathcal{Q}(L)\) be the set of all quantizers for support set of size \(L\).
	For \(\mathcal{U} \subset \mathcal{Q}\), define
	\begin{equation}
		\label{eq:delta-L}
		\begin{gathered}
			\Delta(\mathcal{U}, \lambda) = \min_{q \in \mathcal{U}} C(\lambda, q) - \min_{q \in \mathcal{Q}} C(\lambda, q)\\
			\overline{\Delta}(\mathcal{U}) = \max_{\lambda} \Delta(\mathcal{U}, \lambda) \quad \underline{\Delta}(\mathcal{U}) = \min_{\lambda} \Delta(\mathcal{U}, \lambda).
		\end{gathered}
	\end{equation}
	Here \(\Delta(\mathcal{U}, \lambda)\) is understood as the \(\lambda\)-dependent ``gap to optimality'' when using quantizers from \(\mathcal{U}\) only instead of all quantizers \(\mathcal{Q}\); \(\overline{\Delta}(\mathcal{U})\) and \(\underline{\Delta}(\mathcal{U})\) are (respectively) the worst-case and best-case gap to optimality.
	From these definitions, we have \(0 \leq \underline{\Delta}(\mathcal{U}) \leq \overline{\Delta}(\mathcal{U})\).

\begin{figure*}
		\centering
		\includegraphics{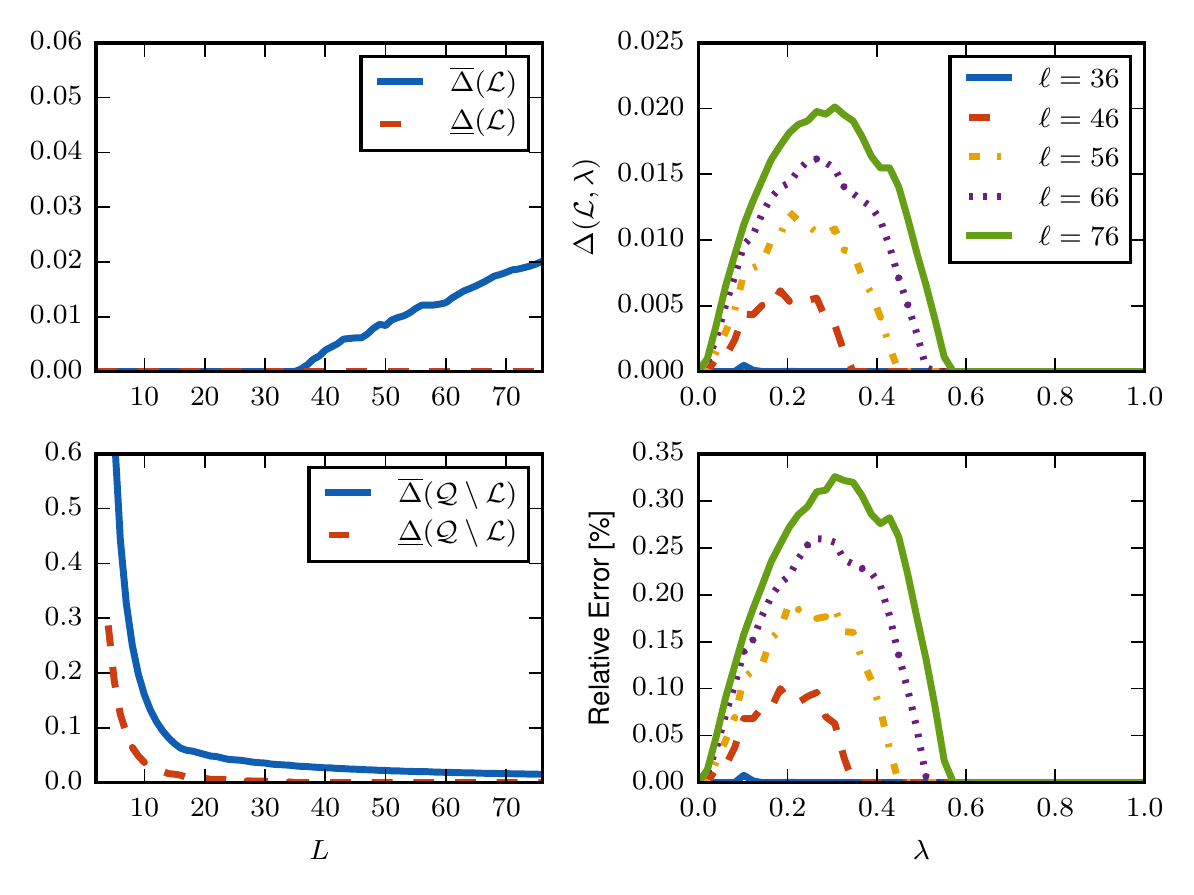}
		\caption{%
		For \(N = 2\):
		\emph{(left)} \(\overline{\Delta}(\cdot)\) and \(\underline{\Delta}(\cdot)\) for \(\mathcal{L}\) \& \(\mathcal{Q} \setminus \mathcal{L}\). %
		\(\overline{\Delta}(\mathcal{L}) > 0\) \& \(\underline{\Delta}(\mathcal{Q} \setminus \mathcal{L}) = 0\) for \(L \geq 35\).
		\emph{(right)} \(\Delta(\mathcal{L}, \lambda)\) \(L \in \{36, 46, 56, 66, 76\}\): %
		\emph{(top)} absolute error and \emph{(bottom)} relative error.
		Max.\ relative error for \(L = 76\) is \(\approx 0.33\%\).
		}
		\label{fig:delta-L}
	\end{figure*}

	The top left side of \fig{delta-L} is a plot of \(\overline{\Delta}(\mathcal{L}(L))\) and \(\underline{\Delta}(\mathcal{\mathcal{L}(L)})\) and the bottom left side is a plot of \(\overline{\Delta}(\mathcal{Q}(L) \setminus \mathcal{L}(L))\) and \(\underline{\Delta}(\mathcal{Q}(L) \setminus \mathcal{L}(L))\) as a function of \(L\).
	When \(N = 2\) and \(L \leq 34\), \(\overline{\Delta}(\mathcal{L}) = 0\) and \(\underline{\Delta}(\mathcal{Q} \setminus \mathcal{L}) > 0\).
	This means that for all \(\lambda\), the optimal quantizer for \(N = 2\), \(L\) can be taken from \(\mathcal{L}\) and there is no value of \(\lambda\) for which the optimal quantizer \emph{can be} taken from \(\mathcal{Q} \setminus \mathcal{L}\).
	These two inequalities imply that \(\mathcal{L}\) is necessary and sufficient for minimizing the cost.
	When \(N = 2\) and \(35 \leq L \leq 76\), from the figure \(0 = \underline{\Delta}(\mathcal{L}) < \overline{\Delta}(\mathcal{L})\) and \(0 = \underline{\Delta}(\mathcal{Q} \setminus \mathcal{L}) < \overline{\Delta}(\mathcal{Q} \setminus \mathcal{L})\).
	This means that there exists a value of \(\lambda\) such that the optimal quantizer \(n^* \not\in \mathcal{L}\).
	The right side of \fig{delta-L} plots the gap to optimality for \(\mathcal{L}(L)\) as a function of \(\lambda\) for representative values of \(L\).
	The top right of the figure plots the absolute magnitude of this gap, whereas the bottom right shows the magnitude of this gap as a relative percentage of the optimal value.
	The worst-case gap is growing in \(L\) (as \fig{delta-L} shows as well), but the gap is still very small, at less than \(0.35\%\), when \(L = 76\).
	There is a range of \(\lambda > 0.6\) for which gap to optimality is identically zero---this range is where the delay component of cost is weighted more heavily than the rate component.
	A minimum delay of 1 can be achieved with the quantizer \((1, \ldots, 1)\) which is in \(\mathcal{L}(L)\).
	Finally, for \(L_1 < L_2\) the interval of \(\lambda\) for which \(\mathcal{L}(L_1)\) is not sufficient is a subset of the interval for which \(\mathcal{L}(L_2)\) is not sufficient.

	The previous counterexample was for the case of \(N = 2\) and we needed \(L \geq 35\) for \(\mathcal{L}(L)\) to no longer be necessary and sufficient for optimally solving \eq{cost-dynamic-program}.
	When \(N\) gets larger, the value of \(L\) at which \(\mathcal{L}(L)\) is no longer necessary and sufficient gets larger as well.

	In summary, when computing the \(\max\) interactively we know that \(\mathcal{L}(L)\):
	\begin{enumerate}
		\item is not an optimal search space for solving \eq{cost-dynamic-program} in general (cf.\ \eq{delta-L} \& \fig{delta-L});
		\item is asymptotically (\(L\) fixed, \(N\) increasing) sufficient for solving \eq{cost-dynamic-program} (cf.\ \propref{asymptotically-optimal});
		\item has \emph{linear} growth (vs.\ exponential for all quantizers) (cf.\ \eq{max-search-family}), and;
		\item incurs a small decrease in performance when it is not optimal (cf.\ \fig{delta-L}).
	\end{enumerate}
	As we show in \secref{results}, adding the binary-search quantizers to \(\mathcal{L}(L)\) results in a simplified search space for computing the \(\argmax\) with little to no incurred penalty.
	Depending upon the system where this interactive quantization strategy is employed, the large reduction in computation costs may more than make up for the small increase in cost that occurs when using \(\mathcal{L}(L)\) for selecting quantizers.

	\section{Results}
	\label{sec:results}
	In this section, we investigate the rate-delay trade-offs, both for the optimized scalar quantizer scheme (\secref{dyn-prog}) and the proposed heuristics (\secref{simple-schemes}).
	For brevity, when we refer to optimal rate-delay trade-offs, we are referring to the rate-delay trade-offs of the optimized scalar quantizer scheme.
	We begin by considering the optimal rate-delay trade-offs for a collection of representative distributions and show that the uniform distribution represents a worst-case distribution.
	This makes sense as the uniform distribution is entropy-maximizing for a given support set size.
	We then investigate the rate-delay trade-offs for binary and max search, demonstrating that these schemes can closely approximate the optimal rate-delay trade-off.

	\subsection{Optimized interactive quantization rate-delay trade-offs}
	\label{sec:results-distributions}
\begin{figure*}
		\centering
		\includegraphics{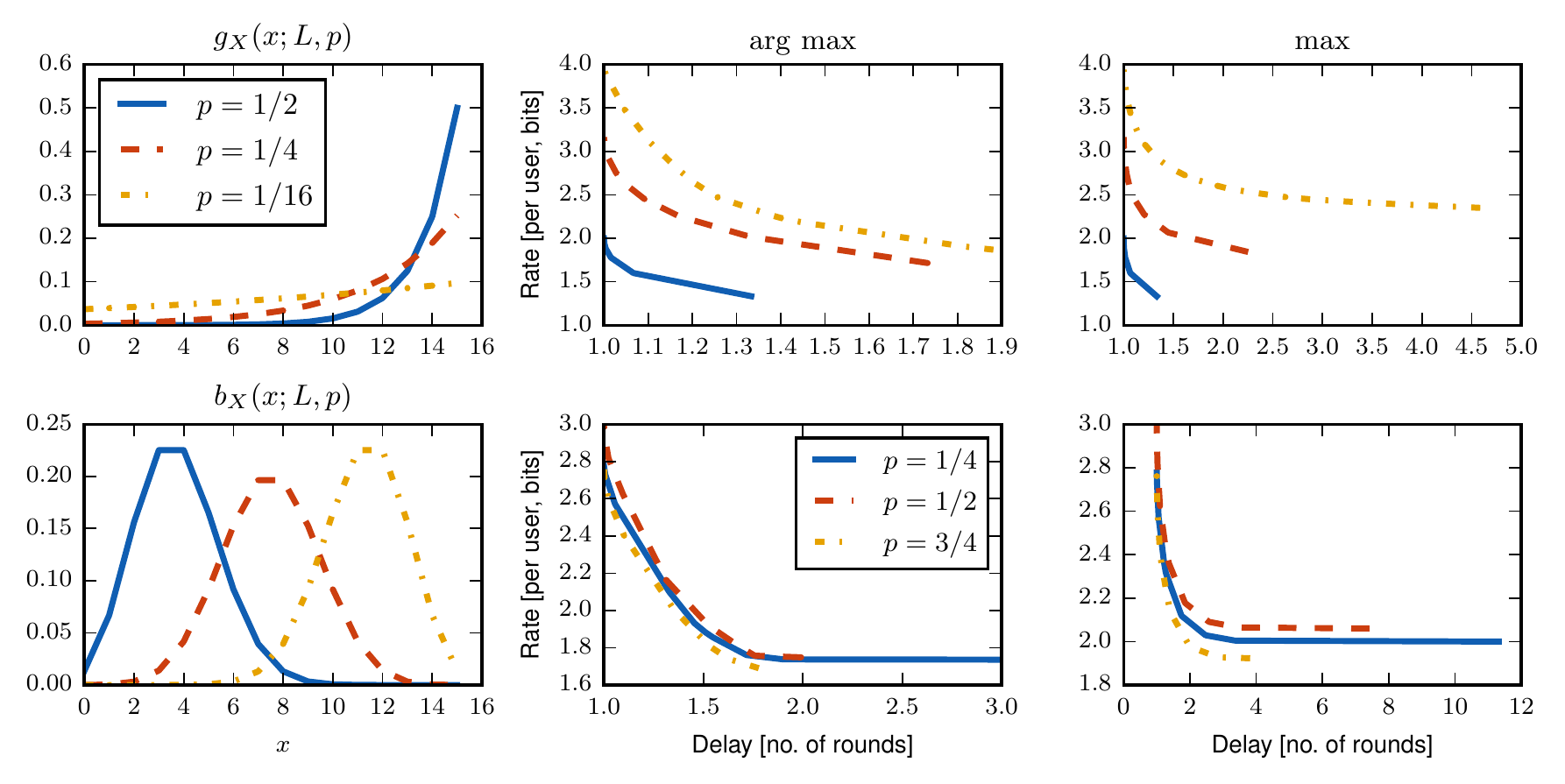}
		\caption{Rate-delay trade-offs for \(g_X(x;L, p)\) (top row) and \(b_X(x; L, p)\) (bottom row) for various values of \(p\). PMFs are shown on the left, rate-delay trade-offs for \(\argmax\) in the center, and \(\max\) on the right.}
		\label{fig:parameterized-distributions}
	\end{figure*}

	For a comparison of the rate-delay trade-offs for various distributions, we consider the following representative distributions parameterized with \(L\) and \(p\) which effects the concentration of the distribution.
	\begin{equation}
		g_X(x; L, p) = \frac{(1-p)^{L - x - 1} p}{1 - (1 - p)^L}, \quad x = 0, \ldots, L - 1
	\end{equation}
	and,
	\begin{equation}
		b_X(x; L, p) = \binom{L}{x}p^x(1-p)^{L - x}, \quad x = 0, \ldots, L - 1.
	\end{equation}
	The distribution \(g_X(x; L, p)\) is shown in \fig{parameterized-distributions} (top left); the effect of varying \(p\) is to vary the ``distance'' from a uniform distribution.
	Also plotted in \fig{parameterized-distributions} is the optimal rate-delay trade-offs for both the \(\argmax\) (top center) and the \(\max\) (top right) functions.
	As \(p\) is made smaller, the trade-offs get worse in that a larger delay is incurred for smaller rates.
	The distribution \(b_X(x; L, p)\) is shown in \fig{parameterized-distributions} (bottom left); unlike \(g_X(x; L, p)\) the ``spread'' of the distribution is not sensitive to the parameter \(p\).
	Looking at the rate-delay trade-offs (bottom center \& left), \(p\) has little effect on the performance of the optimized scheme.

\begin{figure*}
		\centering
		\includegraphics{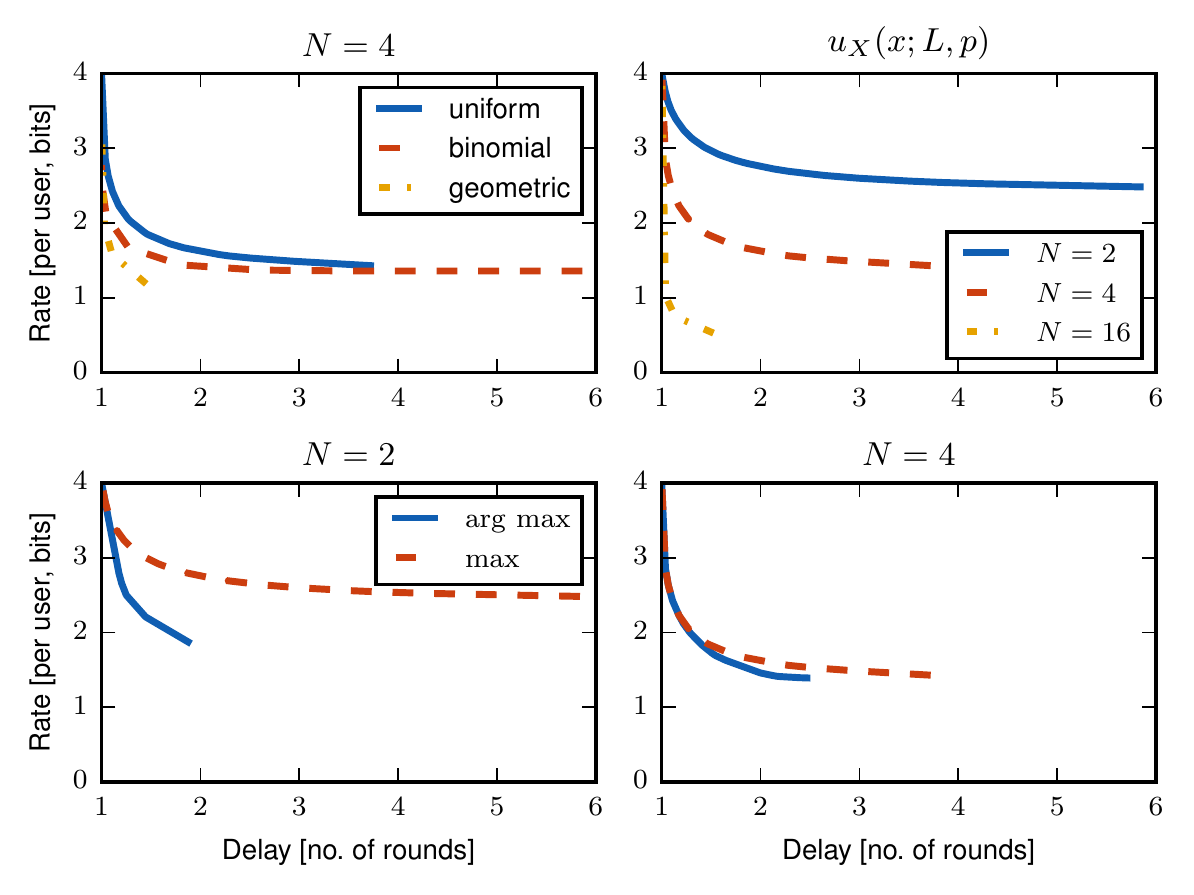}
		\caption{%
		Comparison of rate-delay trade-offs: %
		\emph{(top, left)} for uniform, \(g_X(x; L, p)\), and \(b_X(x; L, p)\) (\(L = 16\)), %
		\emph{(top, right)} when computing the \(\max\) for varying number of users (\(N\)), and %
		\emph{(bottom)} when computing \(\argmax\) vs.\ \(\max\) for \(N = 2\) (right) and \(N = 4\). %
		The source distribution was uniform with support set size \(L = 16\).
		}
		\label{fig:optimal-16-levels}
	\end{figure*}

	\fig{optimal-16-levels} (top left) shows the rate-delay trade-offs for these distributions for both \(\argmax\) and \(\max\) as computed by solving \eq{cost-dynamic-program} and finding the optimal homogeneous quantizer at each round.
	The trade-offs for uniform are worse than for the other distributions.
	For a given upper limit on delay, uniformly distributed sources will require more rate than the other two distributions.
	For a fixed alphabet size, as the number of users is increased (upper right), the trade-offs for the uniform distribution gets better.
	\fig{optimal-16-levels} (bottom row) shows how the rate-delay trade-offs for computing \(\argmax\) and \(\max\) become identical as the number of users increases (cf.\ \propref{max-upper-bound}).
	We see that when \(N\) is small, the CEO is able to compute the \(\argmax\) with either a lower rate (fixed delay) or lower delay (fixed rate) than it would require for computing the \(\max\) with the same fixed rate or delay.
	This difference is especially large in the low rate/high delay regime.
	Doubling the number of users from 2 to 4 significantly reduces this difference.

	\subsection{Extended max search rate-delay trade-offs}
\begin{figure*}
		\centering
		\includegraphics{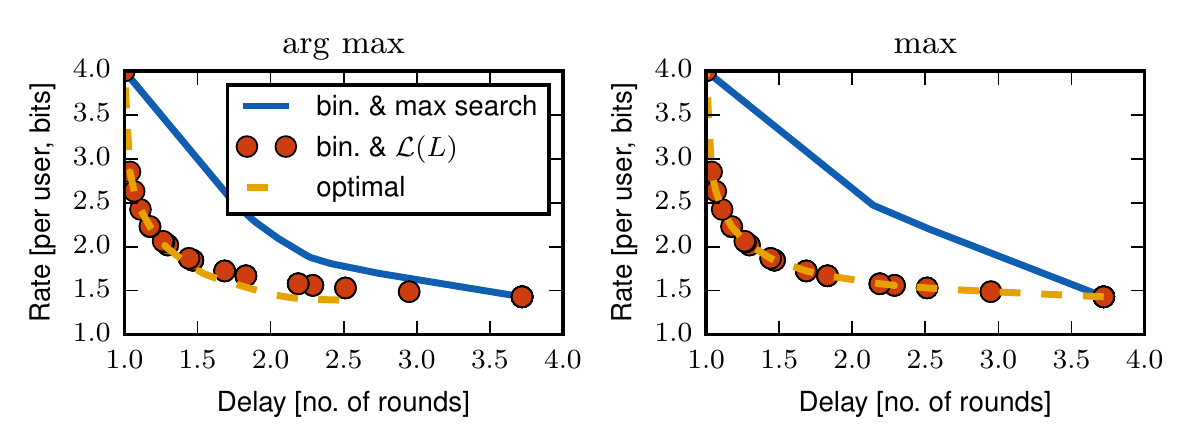}
		\caption{Comparison of rate-delay trade-off for various quantizer search spaces when computing \(\argmax\) (left) and \(\max\) (right). The source distribution was uniform with support set size \(L = 16\) and the number of users was \(N = 4\).}
		\label{fig:simple-16-levels}
	\end{figure*}

	As noted in \secref{simple-schemes}, as the size of the support set \(L\) increases, the number of possible quantizers gets large quickly.
	Based on the simple quantization strategies of binary search and max search, we proposed the extended max search family \(\mathcal{L}(L)\) of quantizers.
	\fig{simple-16-levels} shows the rate-delay trade-off when the search space of \eq{cost-dynamic-program} is taken to be
	\begin{inparaenum}
		\item binary \& max search with the entropy coding quantizer \((1, \ldots, 1)\) (solid line);
		\item binary \& extended max search (which includes the entropy coding quantizer) (circle markers), and;
		\item all possible quantizers (dashed line).
	\end{inparaenum}
	The left side is for the case of computing the \(\argmax\) and the right side is for the case of computing the \(\max\).
	Binary \& max search together can achieve the minimum rate and minimum delay ends of the trade-off curve for both functions, but performs poorly in efficiently trading off delay for rate.
	For computing the \(\argmax\), extended max search is almost equal to the optimal trade-off curve, deviating in the low rate/high delay regime.
	For computing the \(\max\), extended max search equals the optimal trade-off curve.
	By \propref{max-upper-bound} and \fig{optimal-16-levels}, we know that as \(N\) increases, the cost of computing the \(\argmax\) is equal to the cost of computing the \(\max\).
	Even though extended max search is designed with computation of the \(\max\) in mind, \fig{simple-16-levels} and \propref{max-upper-bound} show that it is an effective quantizer search space for computing the \(\argmax\).

	\section{Conclusion}
	\label{sec:conclusions}
	In this paper, we considered the problem of a CEO computing a function of distributed users' state as a model for distributed resource allocation.
	We proposed interactive scalar quantization as an achievable scheme for reducing the required rate to enable the CEO to compute the desired function losslessly at the expense of an increase in delay.
	We solved for optimal rate-delay trade-off of scalar quantization via a dynamic program.
	We established that asymptotically (in the number of users \(N\)), the cost to compute \(\argmax\) is the same as the cost to compute \(\max\).
	By considering simple quantization schemes based on binary search and max search, we designed a family of quantization strategies that is nearly optimal with a significantly reduced computational cost.

	In the present work, we assumed that every user was using the same quantizer (i.e.\ \emph{homogeneous quantization}).
	We know that in the rate-distortion problem, \emph{heterogeneous quantization} can achieve a lower rate for a given distortion than homogeneous quantization \cite{BoyWalWeb2014,RenBoyKu2014}.
	A future direction for the present work would be to extend the model to incorporate different quantizers at the different users.
	Small numerical experiments have demonstrated that substantial further reduction in the rate required to calculate the extremum at a bounded expected delay can be obtained by switching from homogeneous to heterogeneous designs.
	A potential obstacle is the dramatic increase in the size of the search space; the size of the search space is equal to the size of the search space of homogeneous quantization raised to \(N\).

	The current work demonstrates that the required rate can be reduced by tolerating a small increase in delay; in a similar manner, the required rate can be reduced by tolerating a small increase in distortion \cite{BoyWalWeb2014,RenBoyKu2014}.
	The model of these two lines of inquiry could be combined into a single framework to quantify the rate savings that could be realized by tolerating both delay and distortion.
	This would require suitably modifying the cost function of \eq{cost-dynamic-program} to include a term for distortion.
	At each round, the CEO would decide if the distortion is low enough to stop or if communication should continue.

	\IEEEtriggeratref{29}
	\bibliography{IEEEabrv,references}

	\section*{Acknowledgment}
	The authors would like to thank the anonymous reviewers for their valuable comments and suggestions that helped us improve the paper.

	\section*{Disclaimer}
	{\footnotesize The views and conclusions contained herein are those of the authors and should not be interpreted as necessarily representing the official policies or endorsements, either expressed or implied, of the Air Force Research Laboratory or the U.S.\ Government.}
\end{document}